\errorstopmode
\input amssym.def
\input amssym.tex


\magnification=\magstephalf
\hsize=14.0 true cm
\vsize=19 true cm
\hoffset=1.0 true cm
\voffset=2.0 true cm

\abovedisplayskip=12pt plus 3pt minus 3pt
\belowdisplayskip=12pt plus 3pt minus 3pt
\parindent=1.0em


\font\sixrm=cmr6
\font\eightrm=cmr8
\font\ninerm=cmr9

\font\sixi=cmmi6
\font\eighti=cmmi8
\font\ninei=cmmi9

\font\sixsy=cmsy6
\font\eightsy=cmsy8
\font\ninesy=cmsy9

\font\sixbf=cmbx6
\font\eightbf=cmbx8
\font\ninebf=cmbx9

\font\eightit=cmti8
\font\nineit=cmti9

\font\eightsl=cmsl8
\font\ninesl=cmsl9

\font\sixss=cmss8 at 8 true pt
\font\sevenss=cmss9 at 9 true pt
\font\eightss=cmss8
\font\niness=cmss9
\font\tenss=cmss10

\font\eighttt=cmtt8
\font\ninett=cmtt9
\font\tentt=cmtt10

\font\fivemib=cmmib5
\font\sixmib=cmmib6
\font\sevenmib=cmmib7
\font\eightmib=cmmib8
\font\ninemib=cmmib9
\font\tenmib=cmmib10

\font\fivesyb=cmbsy5
\font\sevensyb=cmbsy7
\font\tensyb=cmbsy10

 at 12 true pt
 at 12 true pt
\font\bigrm=cmr10 at 12 true pt
 at 12 true pt
 at 12 true pt

 at 16 true pt
 at 16 true pt
 at 16 true pt
 at 16 true pt
 at 16 true pt

\catcode`@=11
\newfam\mibfam
\newfam\ssfam

\def\tenpoint{\def\rm{\fam0\tenrm}%
    \textfont0=\tenrm \scriptfont0=\sevenrm \scriptscriptfont0=\fiverm
    \textfont1=\teni  \scriptfont1=\seveni  \scriptscriptfont1=\fivei
    \textfont2=\tensy \scriptfont2=\sevensy \scriptscriptfont2=\fivesy
    \textfont3=\tenex \scriptfont3=\tenex   \scriptscriptfont3=\tenex
    \textfont\itfam=\tenit                  \def\it{\fam\itfam\tenit}%
    \textfont\slfam=\tensl                  \def\sl{\fam\slfam\tensl}%
    \textfont\bffam=\tenbf \scriptfont\bffam=\sevenbf
                           \scriptscriptfont\bffam=\fivebf
                           \def\bf{\fam\bffam\tenbf}%
    \textfont\ssfam=\tenss \scriptfont\ssfam=\sevenss
                           \scriptscriptfont\ssfam=\sevenss
                           \def\ss{\fam\ssfam\tenss}%
    \textfont\ttfam=\tentt \def\tt{\fam\ttfam\tentt}%
    \textfont\mibfam=\tenmib \scriptfont\mibfam=\sevenmib
                             \scriptscriptfont\mibfam=\sevenmib
                             \def\mib{\fam\mibfam\tenmib}%
    \normalbaselineskip=13pt
    \setbox\strutbox=\hbox{\vrule height8.5pt depth3.5pt width0pt}%
    \let\big=\tenbig
    \normalbaselines\rm}

\def\ninepoint{\def\rm{\fam0\ninerm}%
    \textfont0=\ninerm      \scriptfont0=\sixrm
                            \scriptscriptfont0=\fiverm
    \textfont1=\ninei       \scriptfont1=\sixi
                            \scriptscriptfont1=\fivei
    \textfont2=\ninesy      \scriptfont2=\sixsy
                            \scriptscriptfont2=\fivesy
    \textfont3=\tenex       \scriptfont3=\tenex
                            \scriptscriptfont3=\tenex
    \textfont\itfam=\nineit \def\it{\fam\itfam\nineit}%
    \textfont\slfam=\ninesl \def\sl{\fam\slfam\ninesl}%
    \textfont\bffam=\ninebf \scriptfont\bffam=\sixbf
                            \scriptscriptfont\bffam=\fivebf
                            \def\bf{\fam\bffam\ninebf}%
    \textfont\ssfam=\niness \scriptfont\ssfam=\sixss
                            \scriptscriptfont\ssfam=\sixss
                            \def\ss{\fam\ssfam\niness}%
    \textfont\ttfam=\ninett \def\tt{\fam\ttfam\ninett}%
    \textfont\mibfam=\ninemib \scriptfont\mibfam=\sixmib
                            \scriptscriptfont\mibfam=\sixmib
                            \def\mib{\fam\mibfam\ninemib}%
    \normalbaselineskip=12pt
    \setbox\strutbox=\hbox{\vrule height8.0pt depth3.0pt width0pt}%
    \let\big=\ninebig
    \normalbaselines\rm}

\def\eightpoint{\def\rm{\fam0\eightrm}%
    \textfont0=\eightrm      \scriptfont0=\sixrm
                             \scriptscriptfont0=\fiverm
    \textfont1=\eighti       \scriptfont1=\sixi
                             \scriptscriptfont1=\fivei
    \textfont2=\eightsy      \scriptfont2=\sixsy
                             \scriptscriptfont2=\fivesy
    \textfont3=\tenex        \scriptfont3=\tenex
                             \scriptscriptfont3=\tenex
    \textfont\itfam=\eightit \def\it{\fam\itfam\eightit}%
    \textfont\slfam=\eightsl \def\sl{\fam\slfam\eightsl}%
    \textfont\bffam=\eightbf \scriptfont\bffam=\sixbf
                             \scriptscriptfont\bffam=\fivebf
                             \def\bf{\fam\bffam\eightbf}%
    \textfont\ssfam=\eightss \scriptfont\ssfam=\sixss
                             \scriptscriptfont\ssfam=\sixss
                             \def\ss{\fam\ssfam\eightss}%
    \textfont\ttfam=\eighttt \def\tt{\fam\ttfam\eighttt}%
    \textfont\mibfam=\eightmib \scriptfont\mibfam=\sixmib
                             \scriptscriptfont\mibfam=\sixmib
                             \def\mib{\fam\mibfam\eightmib}%
    \normalbaselineskip=10pt
    \setbox\strutbox=\hbox{\vrule height7.0pt depth2.0pt width0pt}%
    \let\big=\eightbig
    \normalbaselines\rm}

\def\tenbig#1{{\hbox{$\left#1\vbox to8.5pt{}\right.\n@space$}}}
\def\ninebig#1{{\hbox{$\textfont0=\tenrm\textfont2=\tensy
                       \left#1\vbox to7.25pt{}\right.\n@space$}}}
\def\eightbig#1{{\hbox{$\textfont0=\ninerm\textfont2=\ninesy
                       \left#1\vbox to6.5pt{}\right.\n@space$}}}

\def\setmathbold{
\textfont1=\tenmib \scriptfont1=\sevenmib \scriptscriptfont1=\fivemib
\textfont2=\tensyb \scriptfont2=\sevensyb \scriptscriptfont2=\fivesyb}

\def\unsetmathbold{
\textfont1=\teni \scriptfont1=\seveni \scriptscriptfont1=\fivei
\textfont2=\tensy \scriptfont2=\sevensy \scriptscriptfont2=\fivesy
}

\font\sectionfont=cmbx10
\font\subsectionfont=cmti10

\def\figurecaptionfont{\ninepoint}
\def\tablecaptionfont{\ninepoint}
\def\footnotefont{\eightpoint}


\newcount\equationno
\newcount\bibitemno
\newcount\figureno
\newcount\tableno

\equationno=0
\bibitemno=0
\figureno=0
\tableno=0


\footline={\ifnum\pageno=0{\hfil}\else
{\hss\rm\the\pageno\hss}\fi}


\def\section #1. #2 \par
{\vskip0pt plus .10\vsize\penalty-100 \vskip0pt plus-.10\vsize
\vskip 1.6 true cm plus 0.2 true cm minus 0.2 true cm
\global\def\equationlabel{#1}
\global\equationno=0
\setmathbold
\leftline{\sectionfont #1. #2}\par
\immediate\write\terminal{Section #1. #2}\unsetmathbold
\vskip 0.7 true cm plus 0.1 true cm minus 0.1 true cm
\noindent}


\def\subsection #1 \par
{\vskip0pt plus 1.0 true cm\penalty-50 \vskip0pt plus-1.0 true cm
\vskip2.5ex plus 0.1ex minus 0.1ex
\leftline{\subsectionfont #1}\par
\immediate\write\terminal{Subsection #1}
\vskip1.0ex plus 0.1ex minus 0.1ex
\noindent}


\def\appendix #1. #2 \par
{\vskip0pt plus .10\vsize\penalty-100 \vskip0pt plus-.10\vsize
\vskip 1.6 true cm plus 0.2 true cm minus 0.2 true cm
\global\def\equationlabel{\hbox{\rm#1}}
\global\equationno=0
\setmathbold
\leftline{\sectionfont Appendix #1. #2}\par
\immediate\write\terminal{Appendix #1. #2}\unsetmathbold
\vskip 0.7 true cm plus 0.1 true cm minus 0.1 true cm
\noindent}



\def\equation#1{$$\displaylines{\qquad #1}$$}
\def\enum{\global\advance\equationno by 1
\hfill\llap{{\rm(\equationlabel.\the\equationno)}}}
\def\noenum{\hfill}

\def\nexteq#1{\cr\noalign{\vskip#1}\qquad}


\def\ifundefined#1{\expandafter\ifx\csname#1\endcsname\relax}

\def\ref#1{\ifundefined{#1}?\immediate\write\terminal{unknown reference
on page \the\pageno}\else\csname#1\endcsname\fi}

\newwrite\terminal
\newwrite\bibitemlist

\def\bibitem#1#2\par{\global\advance\bibitemno by 1
\immediate\write\bibitemlist{\string\def
\expandafter\string\csname#1\endcsname
{\the\bibitemno}}
\item{[\the\bibitemno]}#2\par}

\def\beginbibliography{
\vskip0pt plus .15\vsize\penalty-100 \vskip0pt plus-.15\vsize
\vskip 1.2 true cm plus 0.2 true cm minus 0.2 true cm
\leftline{\sectionfont References}\par
\immediate\write\terminal{References}
\immediate\openout\bibitemlist=biblist
\frenchspacing\parindent=1.8em
\vskip 0.5 true cm plus 0.1 true cm minus 0.1 true cm}

\def\endbibliography{
\immediate\closeout\bibitemlist
\nonfrenchspacing\parindent=1.0em}


\def\figurecaption#1{\global\advance\figureno by 1
\narrower\figurecaptionfont Fig.~\the\figureno. #1}

\def\tablecaption#1{\global\advance\tableno by 1
\centerline{\tablecaptionfont Table~\the\tableno. #1}}

\def\thicktablerule{\hrule height0.8pt}
\def\thintablerule{\hrule height0.4pt}

\tenpoint

\immediate\openin\bibitemlist=biblist
\ifeof\bibitemlist\immediate\closein\bibitemlist
\else\immediate\closein\bibitemlist

 \fi


\def\thismonth{\ifcase\month\or
January\or February\or March\or April\or May\or June\or
July\or August\or September\or October\or November\or December\fi}

\font\fourteenrm=cmr12 scaled \magstep1
\font\fourteeni=cmmi12 scaled \magstep1
\font\fourteensy=cmsy10 scaled \magstep2
\font\fourteenex=cmex10 scaled \magstep2
\font\fourteenbf=cmbx12 scaled \magstep1
\font\fourteensl=cmsl12 scaled \magstep1
\font\fourteentt=cmtt12 scaled \magstep1
\font\fourteenit=cmti12 scaled \magstep1

\ifx\fourteenpoint\undefined
   \def\fourteenpoint{\def\rm{\fam0\fourteenrm}
       \textfont0=\fourteenrm \scriptfont0=\tenrm \scriptscriptfont0=\sevenrm
       \textfont1=\fourteeni  \scriptfont1=\teni  \scriptscriptfont1=\seveni
       \textfont2=\fourteensy \scriptfont2=\tensy \scriptscriptfont2=\sevensy
       \textfont3=\fourteenex \scriptfont3=\fourteenex
                              \scriptscriptfont3=\fourteenex
       \textfont\itfam=\fourteenit  \def\it{\fam\itfam\fourteenit}%
       \textfont\slfam=\fourteensl  \def\sl{\fam\slfam\fourteensl}%
       \textfont\ttfam=\fourteentt  \def\tt{\fam\ttfam\fourteentt}%
       \textfont\bffam=\fourteenbf  \scriptfont\bffam=\tenbf
        \scriptscriptfont\bffam=\sevenbf  \def\bf{\fam\bffam\fourteenbf}%
       \normalbaselineskip=17pt
       \setbox\strutbox=\hbox{\vrule height11.9pt depth6.3pt width0pt}%
       \normalbaselines\rm}
   \fi

\font\seventeenrm=cmr17
\font\seventeeni=cmmi12 scaled \magstep2
\font\seventeensy=cmsy10 scaled \magstep3
\font\seventeenex=cmex10 scaled \magstep3
\font\seventeenbf=cmbx12 scaled \magstep2
\font\seventeensl=cmsl12 scaled \magstep2
\font\seventeentt=cmtt12 scaled \magstep2
\font\seventeenit=cmti12 scaled \magstep2

\font\twelverm=cmr12
\font\twelvei=cmmi12
\font\twelvesy=cmsy10 scaled \magstep1
\font\twelvebf=cmbx12

\ifx\seventeenpoint\undefined
   \def\seventeenpoint{\def\rm{\fam0\seventeenrm}
       \textfont0=\seventeenrm \scriptfont0=\twelverm \scriptscriptfont0=\ninerm
       \textfont1=\seventeeni  \scriptfont1=\twelvei  \scriptscriptfont1=\ninei
       \textfont2=\seventeensy \scriptfont2=\twelvesy \scriptscriptfont2=\ninesy
       \textfont3=\seventeenex \scriptfont3=\seventeenex
                              \scriptscriptfont3=\seventeenex
       \textfont\itfam=\seventeenit  \def\it{\fam\itfam\seventeenit}%
       \textfont\slfam=\seventeensl  \def\sl{\fam\slfam\seventeensl}%
       \textfont\ttfam=\seventeentt  \def\tt{\fam\ttfam\seventeentt}%
       \textfont\bffam=\seventeenbf  \scriptfont\bffam=\twelvebf
        \scriptscriptfont\bffam=\ninebf  \def\bf{\fam\bffam\seventeenbf}%
       \normalbaselineskip=21pt
       \setbox\strutbox=\hbox{\vrule height17pt depth4pt width0pt}%
       \normalbaselines\rm}
   \fi

\input epsf
\epsfclipon



\def\rmd{{\rm d}}

\def\rmO{{\rm O}}


\def\Re{\kern1pt{\rm Re}\,}


\def\proof{\noindent{\sl Proof:}\kern0.6em}

\def\frac#1#2{\hbox{$#1\over#2$}}
\def\dual{\kern0.1em\mathstrut^*\kern-0.2em}

\def\lvec#1{\setbox0=\hbox{$#1$}
    \setbox1=\hbox{$\scriptstyle\leftarrow$}
    #1\kern-\wd0\smash{
    \raise\ht0\hbox{$\raise1pt\hbox{$\scriptstyle\leftarrow$}$}}
    \kern-\wd1\kern\wd0}
\def\rvec#1{\setbox0=\hbox{$#1$}
    \setbox1=\hbox{$\scriptstyle\rightarrow$}
    #1\kern-\wd0\smash{
    \raise\ht0\hbox{$\raise1pt\hbox{$\scriptstyle\rightarrow$}$}}
    \kern-\wd1\kern\wd0}

\def\slash#1{\setbox2=\hbox{$\displaystyle#1$}%
             \setbox3=\hbox{$\displaystyle/$}%
             #1\kern-0.8\wd2/\kern-1.0\wd3\kern0.8\wd2\kern0.5pt}


\def\nabstar#1{\nabla\kern-0.5pt\smash{\raise 4.5pt\hbox{$\ast$}}
               \kern-4.5pt_{#1}}

\def\drvstar#1{\partial\kern-0.5pt\smash{\raise 4.5pt\hbox{$\ast$}}
               \kern-5.0pt_{#1}}




\def\psibar{\overline{\psi}}

\def\Psibar{\overline{\Psi\kern-0.06em}\kern0.06em}
\def\cbar{\bar{c}}

\def\etabar{\bar{\eta}}

\def\Cbar{\overline{C\kern-0.05em}\kern0.05em}
\def\Mbar{\kern0.15em\overline{\kern-0.15em M\kern-0.05em}\kern0.05em}


\def\dirac#1{\gamma_{#1}}
\def\diracstar#1#2{
    \setbox0=\hbox{$\gamma$}\setbox1=\hbox{$\gamma_{#1}$}
    \gamma_{#1}\kern-\wd1\kern\wd0
    \smash{\raise4.5pt\hbox{$\scriptstyle#2$}}}


\def\SUn{{\rm SU}(N)}

\def\tr{\hbox{\rm tr}}

\def\Ad{{\rm Ad}\,}




\def\mbar{\kern1pt\overline{\kern-1pt m\kern-1pt}\kern1pt}
\def\mb#1{m_{0,#1}}

\def\Rmat{{\cal R}}
\def\Xmat{{\cal X}}

\def\msbar{{\rm \overline{MS\kern-0.05em}\kern0.05em}}

\def\As#1{A^s_{#1}}

\def\FFR{(F\dual F)_{\hbox{\sixrm R}}}
\def\AsR#1{(A^s_{#1})_{\hbox{\sixrm R}}}
\def\AsRp#1{(A^s_{#1})_{\hbox{\sixrm R'}}}
\def\mPR{(mP)_{\hbox{\sixrm R}}}

\def\FFRGI{(F\dual F)_{\hbox{\sixrm RGI}}}
\def\AsRGI#1{(A^s_{#1})_{\hbox{\sixrm RGI}}}
\def\mPRGI{(mP)_{\hbox{\sixrm RGI}}}

\def\OR#1{({\cal O}_{#1})_{\hbox{\sixrm R}}}
\def\ORGI#1{({\cal O}_{#1})_{\hbox{\sixrm RGI}}}



\def\Stot{S_{\rm tot}}


\def\eps{\epsilon}
\def\obs{{\cal O}}
\def\Nf{N_{\rm f}}
\def\dx#1{{\rm d}x_{#1}}
\def\dbrs{\delta_{\scriptscriptstyle\rm BRS}}
\def\dbrsc{\hat\delta_{\scriptscriptstyle\rm BRS}}
\def\dshf{\delta_{\rm s}}
\def\dB{\rmd^B\kern-1pt}
\def\Gzero{\Gamma^{(0)}}
\def\Gone{\Gamma^{(1)}}
\def\GR{\Gamma_R}

\def\GRone{\Gamma^{(1)}_R}
\def\GRn#1{\Gamma_{R,#1}}

\def\GRonen#1{\Gamma^{(1)}_{R,#1}}

%
\rightline{CERN-TH-2021-041,  MPP-2021-38}

\vskip1.2cm
{\fourteenpoint
\centerline{Renormalization of the topological charge density}
\vskip 0.2 true cm
\centerline{ in QCD with dimensional regularization}
}

\tenpoint
\vskip 0.6 true cm
\centerline{\bigrm Martin L\"uscher$^{\rm a,b}$ and Peter Weisz$^{\rm c}$}

\vskip1.5ex
\centerline{{\it $^{\rm a}$CERN,
Theoretical Physics Department, 1211 Geneva 23, Switzerland}}

\vskip1.0ex
\centerline{{\it $^{\rm b}$Albert Einstein Center for Fundamental Physics}}
\centerline{{\it Institute for Theoretical Physics,
Sidlerstrasse 5, 3012 Bern, Switzerland}}

\vskip1.0ex
\centerline{{\it $^{\rm c}$Max-Planck-Institut f\"ur Physik}}
\centerline{{\it F\"ohringer Ring 6, 80805 Munich, Germany}}

\vskip 0.8 true cm
\thintablerule
\vskip 2.0ex
\ninepoint
\noindent
To all orders of perturbation theory,
the renormalization of the topological charge density
in dimensionally regularized QCD is shown to require no more than
an additive renormali\-zation proportional to the divergence of
the flavour-singlet axial current.
The proof is based on the standard BRS analysis of the QCD
vertex functional in the background gauge and exploits
the special algebraic properties of the charge density
through the Stora--Zumino chain of descent equations.

\vskip 2.0ex
\thintablerule

\tenpoint


\section 1. Introduction

All known consistent forms of dimensional regularization of QCD
break chiral symmetry and the symmetry is then only recovered
after renormalization and removal of the regularization.
In the flavour-singlet channel, the situation is further
complicated by the chiral anomaly,
a term proportional to the topological charge density
in the axial-current conservation equation, which
requires renormalization as do the other terms in that equation.

Parity-odd fields are in general not easy to deal with in
dimensional regularization, because the fifth Dirac matrix $\dirac{5}$
and the Levi--Civita symbol $\eps_{\mu\nu\rho\sigma}$
are not naturally defined in dimensions other than four.
In QCD this technical difficulty can however be bypassed
by representing such fields through totally
antisymmetric tensor fields [\ref{VermaserenLarin},\ref{Larin}].
Using this representation, the renormalization of the
axial quark densities, the axial currents and the
chiral anomaly
has been worked out to high order in the gauge coupling
[\ref{Larin}--\ref{AhmedII}].

In these computations,
a multiplicative renormalization of the topological charge density
turned out to be unnecessary,
thus suggesting that the density is finite
to all orders up to additive renormalizations
[\ref{AhmedII}].
The absence of a divergent multiplicative renormalization
is perhaps not unexpected in view of the Adler--Bardeen theorem
[\ref{AB}] or simply because the topological charge of classical fields
assumes integer values.
Many years ago, Breitenlohner, Maison and Stelle [\ref{BMS}]
attempted to trace back the finiteness
of the charge density to its
algebraic properties, but their
argumentation remained incomplete and was partially incorrect
to the extent of being inconclusive.
To date a rigorous all-order discussion of the situation
in dimensionally regularized QCD appears to be
missing and
the principal goal here is to fill this gap, using
the standard BRS analysis of the QCD vertex functional
[\ref{BRSI}--\ref{ZJE}],
the background gauge [\ref{DeWittI}--\ref{KlubergZuberII}]
and the Stora--Zumino descent equations [\ref{StoraI}--\ref{Zumino}].

After going through some definitions and preliminary material
in sect.~2,
the rather special algebraic properties of the topological
charge density, as expressed through the descent equations,
are exposed in sect.~3.
The symmetries of the QCD vertex functional in
presence of a background gauge field and
the sources for the descendants of the charge density
are then discussed.
These strongly constrain the form of the divergent parts
of the vertex functional and,
as shown in sect.~5, eventually
exclude a multiplicative renormalization of the
charge density. The
paper ends with some comments on
the axial anomaly
and a few concluding remarks.

\section 2. Preliminaries

\vskip-2.5ex

\subsection 2.1 Background field technique

The theory is set up in Euclidean space in the standard manner
with any number $\Nf\geq0$ of quarks
in the fundamental representation
of the gauge group $\SUn$ (see appendix A for any unexplained notation).

In the background field formalism,
the fundamental gauge potential $A_{\mu}(x)$
is normalized such that
the associated field tensor $F_{\mu\nu}(x)$ and
the gauge-covariant derivatives do not involve the gauge coupling.
The QCD action in $D$ dimensions is then given by
\equation{
  S=\int\rmd^D\! x\,
  \biggl\{-{1\over2 g_0^2}\tr\{F_{\mu\nu}F_{\mu\nu}\}
  +\sum_{r=1}^{\Nf}\psibar_r(\slash{D}+\mb{r})\psi_r
  \biggr\},
  \enum
}
where the index $r$ of the quark fields $\psibar_r$ and
$\psi_r$ labels the quark flavours,
$g_0$ is the bare coupling and
$\mb{r}$ the bare mass of the quark number $r$.

The background field technique permits the theory
to be probed with greater respect for the gauge symmetry
than is the case when probed in conventional ways
[\ref{DeWittI}--\ref{KlubergZuberII}].
Let $B_{\mu}(x)$ be a smooth classical gauge potential and consider
the decomposition
\equation{
  A_{\mu}(x)=B_{\mu}(x)+g_0q_{\mu}(x)
  \enum
}
of the fundamental gauge potential $A_{\mu}$ in the background field $B_{\mu}$
and the quantum field $q_{\mu}$, which is now the field integrated
over in the functional integral.
A possible choice of the gauge-fixing and associated ghost action
is then
\equation{
  S_{\rm gf}=-\lambda_0\int\rmd^D\!x\,
  \tr\left\{D_{\mu}q_{\mu}D_{\nu}q_{\nu}\right\},
  \qquad
  D_{\mu}=\partial_{\mu}+\Ad B_{\mu},
  \enum
  \nexteq{2.0ex}
  S_{\rm gh}=-2\int\rmd^D\!x\,
  \tr\left\{D_{\mu}\cbar\left(D_{\mu}+g_0\Ad q_{\mu}\right)c\right\},
  \enum
}
$c$ and $\cbar$ being the ghost and antighost fields.
As is quite clear from these expressions, and further
discussed in subsect.~3.2, this way of fixing the gauge
preserves a classical gauge symmetry.

The theory with total action
\equation{
  \Stot=S+S_{\rm gf}+S_{\rm gh}
  \enum
}
has a regular perturbation expansion in Feynman diagrams,
if the background field is treated as
an additional source field, i.e.~if the functional integral
is expanded in a powers series in this field.
In particular, at
vanishing background field, the theory in the standard
Lorentz-covariant gauge is recovered.
All-important is then the fact that the theory
renormalizes in the same way with and without background field,
the latter requiring no renormalization.

An introduction to the subject and a proof of the renormalizability
of the dimensionally regularized theory
in presence of the background field is provided in
the first few sections of ref.~[\ref{BFM}].
Some of the strategies described there will again be used
here, but the presentation in the following is intended to be
self-contained.

\subsection 2.2 Tensor fields and the topological charge density

In $D=4$ dimensions,
the topological charge density is given by
\equation{
  q_{\rm top}(x)=-{1\over32\pi^2}\eps_{\mu\nu\rho\sigma}
  \tr\{F_{\mu\nu}(x)F_{\rho\sigma}(x)\}.
  \enum
}
Since the Levi--Civita symbol $\eps_{\mu\nu\rho\sigma}$
is not a well defined object in dimensions other than four,
the use of eq.~(2.6) in any dimension would
require the first four dimensions to be distinguished from
the $-2\eps$ extra dimensions.

Following refs.~[\ref{VermaserenLarin}--\ref{AhmedII}],
such a distinction can be avoided by
noting that the totally antisymmetric tensor field
\equation{
  (FF)_{\mu\nu\rho\sigma}(x)
  =
  F^a_{\mu\nu}(x)F^a_{\rho\sigma}(x)+
  F^a_{\nu\rho}(x)F^a_{\mu\sigma}(x)+
  F^a_{\nu\sigma}(x)F^a_{\rho\mu}(x)
  \enum
}
is, in four dimensions, proportional to $\eps_{\mu\nu\rho\sigma}$ times
the charge density.
This tensor field is well defined in any dimension and
thus provides a possible representation of the
charge density in the framework of dimensional regularization.

In the case of the axial quark densities and currents,
the totally antisymmetric tensor fields
\equation{
  P^{rs}_{\mu\nu\rho\sigma}(x)=
  \psibar_r(x)
  \dirac{[\mu}\dirac{\vphantom{[}\nu}\dirac{\vphantom{[}\rho}\dirac{\sigma]}
  \psi_s(x),
  \enum
  \nexteq{2.5ex}
  A^{rs}_{\mu\nu\rho}(x)=
  \psibar_r(x)\dirac{[\mu}\dirac{\vphantom{[}\nu}\dirac{\rho]}\psi_s(x),
  \enum
}
may similarly be taken as a possible representation of these
fields in arbitrary dimensions
(following common practice, an antisymmetrization over the indices
enclosed in square brackets is implied).
It may be worth mentioning in passing that
the fields (2.8),(2.9) satisfy an exact PCAC relation,
in any dimension, involving an evanescent further field, which
renormalizes the other fields in the equation
and gives rise to the axial anomaly at $D=4$.

Correlation functions of tensor fields and the
fundamental fields can be worked out in
perturbation theory in the standard manner.
Covariance under the full Lorentz group, including parity,
is exactly preserved in these calculations and the required
counterterms are Lorentz-covariant polynomials
in the external momenta, Kronecker deltas and products of
Dirac matrices (if some external
Dirac indices are uncontracted).
In particular, tensor fields renormalize among themselves.

\section 3. Algebraic properties of the topological charge density

Totally antisymmetric tensors like $(FF)_{\mu\nu\rho\sigma}$
are naturally associated with differential forms.
In this particular case, the form is, in any dimension,
proportional to the second Chern character and thus
has some special algebraic properties.

\subsection 3.1 BRS variation [\ref{BRSI},\ref{BRSII}]

In presence of the background gauge potential $B_{\mu}$, the
BRS variation of the quantum field $q_{\mu}$, the ghost
fields $c$, $\cbar$ and the quark fields $\psi,\psibar$ is given by
\equation{
  \dbrs q_{\mu}=\left(D_{\mu}+g_0\Ad q_{\mu}\right)c,
  \enum
  \nexteq{2.0ex}
  \dbrs c=-g_0c^2,
  \enum
  \nexteq{2.0ex}
  \dbrs\cbar=\lambda_0D_{\mu}q_{\mu},
  \enum
  \nexteq{2.0ex}
  \dbrs\psi=-g_0c\psi,\qquad \dbrs\psibar=-g_0\psibar c.
  \enum
}
Since the
background field is not transformed, eq.~(3.1) implies
\equation{
  \dbrs A_{\mu}=g_0\left(\partial_{\mu}+\Ad A_{\mu}\right)c,
  \enum
}
which shows that the BRS transformation of the gauge potential
$A_{\mu}$ is an infinitesimal gauge transformation.

The BRS variation is an antiderivative with respect to
the grading defined by the fermion (ghost plus quark) number, i.e.
\equation{
  \dbrs(fg)=\dbrs fg+(-1)^nf\dbrs g
  \enum
}
if $f$ has fermion number $n$. When acting on differential forms,
the rank of the form is often included in the grading and the
exterior differential $\rmd$ then anticommutes with $\dbrs$.
In the present context, where the BRS variation eventually
acts on tensor fields, this convention however tends to be confusing
and is not applied.

\subsection 3.2 Background gauge variation

Background gauge transformations are generated by classical
fields $\omega(x)$ with values in the Lie algebra of $\SUn$.
The associated gauge variation includes the background field
\equation{
  \delta_{\omega}B_{\mu}=D_{\mu}\omega
  \enum
}
and acts on the quantum fields according to
\equation{
  \delta_{\omega}q_{\mu}=[q_{\mu},\omega],
  \enum
  \nexteq{2.0ex}
  \delta_{\omega}c=[c,\omega],\qquad
  \delta_{\omega}\cbar=[\cbar,\omega],
  \enum
  \nexteq{2.0ex}
  \delta_{\omega}\psi=-\omega\psi,\qquad
  \delta_{\omega}\psibar=\psibar\omega.
  \enum
}
Gauge and BRS transformations both preserve the total action
(2.5) and their commutator $[\dbrs,\delta_{\omega}]$ vanishes.

\subsection 3.3 Descent equations

The Stora--Zumino chain of equations [\ref{StoraI}--\ref{Zumino}],
which descends from the second Chern character, make the
special algebraic properties of the latter explicit
(for an introduction to the
subject see ref.~[\ref{Bertlmann}], for example).
It is now helpful to introduce the differential forms
\equation{
  A=A_{\mu}\dx{\mu},
  \qquad
  F=\frac{1}{2}F_{\mu\nu}\dx{\mu}\dx{\nu}=
  \rmd A+A^2,
  \enum
}
and similarly the forms $B$ and $q$.
Starting from the 4-form
\equation{
  \tr\{F^2\}=-{1\over4!}(FF)_{\mu\nu\rho\sigma}
  \dx{\mu}\dx{\nu}\dx{\rho}\dx{\sigma},
  \enum
}
a sequence $\phi_3,\phi_2,\phi_1,\phi_0$ of differential
forms of decreasing rank may then be constructed
satisfying the descent equations
\equation{
  \tr\{F^2\}=\rmd\phi_3,
  \enum
  \nexteq{2.0ex}
  \dbrs\phi_k=\rmd\phi_{k-1},\qquad k=3,2,1,
  \enum
  \nexteq{2.0ex}
  \dbrs\phi_0=0.
  \enum
}
The particular solution of these equations chosen here,
\equation{
  \phi_3=
  \tr\bigl\{A\kern1pt\rmd A+\frac{2}{3}A^3\bigr\}
  -g_0\rmd(\tr\{qB\}),
  \enum
  \nexteq{2.0ex}
  \phi_2=
   g_0\kern1pt\tr\{c\kern1pt\rmd A\}
   -g_0\dbrs(\tr\{qB\})
   +g_0\rmd(\tr\{cB\}),
  \enum
  \nexteq{2.0ex}
  \phi_1=
  g_0^2\kern1pt\tr\{c^2A\}
  +g_0\dbrs(\tr\{cB\}),
  \enum
  \nexteq{2.0ex}
  \phi_0=g_0^3\kern1pt\frac{1}{3}\tr\{c^3\},
  \enum
}
includes several terms proportional to the background field $B$,
which serve to ensure a simple transformation behaviour,
\equation{
  \delta_{\omega}\phi_3=\tr\{\rmd\omega\kern1pt\rmd B\},
  \enum
  \nexteq{2.0ex}
  \delta_{\omega}\phi_k=0,\quad k=0,1,2,
  \enum
}
under background gauge transformations.

\section 4. Definition and symmetries of the bare vertex functional

The renormalization of the theory
with insertions of composite fields
will be studied by adding sources for all relevant fields and by
discussing the possible structure of the divergent parts
of the associated vertex functional.
Since the quark fields give rise to only minor complications
in this analysis,
the pure gauge theory will now first be considered,
the modifications required in full QCD being discussed in subsect.~5.6.

\subsection 4.1 Source terms for the basic fields

Following standard practice, the basic source terms
included in the QCD functional integral are
\equation{
  (J,q)+(\etabar,c)+(\cbar,\eta)+(K,\dbrs q)-(L,\dbrs c),
  \enum
}
where $J_{\mu},\etabar,\eta,K_{\mu}$ and $L$ are classical
source fields with values in the Lie algebra of $\SUn$.
Scalar products of such coloured fields like
\equation{
  (J,q)=\int\rmd^D\!x\,J^a_{\mu}(x)q^a_{\mu}(x)
  \enum
}
are defined in the obvious way and it is understood that
$\etabar,\eta$ and $K_{\mu}$ are fermion fields that anticommute
with the ghost fields.

The source terms (4.1) are such that
the application of the BRS variation to the sum of terms
is equivalent to a change of the source fields.
This property is shared by the further source terms introduced below
and eventually ensures that the BRS symmetry
turns into a symmetry of the vertex functional.

\subsection 4.2 Sources for the descendants of the charge density

Appropriate source fields for the totally antisymmetric coefficients
$(\phi_k)_{\mu_1\ldots\mu_k}$ of the differential forms (3.16)--(3.19)
are classical tensor fields $(H_k)_{\mu_1\ldots\mu_k}$ of the same type.
The corresponding source terms,
\equation{
  (H_k,\phi_k)=
  \int\rmd^D\!x\,
  (H_k)_{\mu_1\ldots\mu_k}(x)(\phi_k)_{\mu_1\ldots\mu_k}(x),
  \enum
}
transform under the BRS variation
according to\kern1pt\footnote{$\dagger$}{\footnotefont%
Algebraic consistency requires that
source fields with odd fermion number anticommute with the BRS
variation of the quantum fields.}
\equation{
  \dbrs(H_0,\phi_0)=0,
  \enum
  \nexteq{2.0ex}
  \dbrs(H_k,\phi_k)
  =(-1)^{k-3}(\rmd^{\ast}\kern-1pt H_k,\phi_{k-1}),
  \quad
  k=1,2,3,
  \enum
}
where
\equation{
  (\rmd^{\ast}\kern-1pt H_k)_{\mu_1\ldots\mu_{k-1}}(x)
  =-\partial_{\mu}(H_k)_{\mu\mu_1\ldots\mu_{k-1}}(x).
  \enum
}
As will become clear in sect.~5,
source terms for two further fields,
\equation{
  s_{\mu}=\tr\{cq_{\mu}\}
  \hskip0.5em\hbox{and}\hskip0.5em
  \dbrs s_{\mu},
  \enum
}
must be included together with the terms (4.3)
to be able to renormalize the correlation functions with
insertions of
$(\phi_1)_{\mu}$ and $(\phi_2)_{\mu\nu}$.

\topinsert
\newdimen\digitwidth
\setbox0=\hbox{$0$}
\digitwidth=\wd0
\catcode`@=\active
\def@{\kern\digitwidth}
\newdimen\signwidth
\setbox1=\hbox{$-$}
\signwidth=\wd1
\tablecaption{Properties of the source fields}
\vskip-4.0ex

$$\vbox{\settabs\+&%
                  xxxxxxx&xx&%
                  xxxxxxxxxx&x&%
                  xxxxxxxxxx&xxxx&%
                  xxxxxxx&xx&%
                  xxxxxxxxxx&x&%
                  xxxxxxxxxx&x&\cr
\thicktablerule
\vskip1.2ex
                \+& \hfill Field\hfill
		 && \hfill Dimension\hfill
                 && \hfill Ghost no\hfill
                 && \hfill Field\hfill
		 && \hfill Dimension\hfill
                 && \hfill Ghost no\hfill
                 &\cr
\vskip0.8ex
\thintablerule
\vskip1.2ex
{\ninepoint
  \+& \hskip4ex $J^a_{\mu}$\hfill
  &&  \hfill $3$\hfill
  &&  \hfill $\kern7pt0$\hfill
  &&  \hskip3ex $E_{\mu}$\hfill
  &&  \hfill $2$\hfill
  &&  \hfill $-1$\hfill
  &\cr
  \+& \hskip4ex $\etabar^a$\hfill
  &&  \hfill $3$\hfill
  &&  \hfill $-1$\hfill
  &&  \hskip3ex $F_{\mu}$\hfill
  &&  \hfill $1$\hfill
  &&  \hfill $-2$\hfill
  &\cr
  \+& \hskip4ex $\eta^a$\hfill
  &&  \hfill $3$\hfill
  &&  \hfill $\kern7pt1$\hfill
  &&  \hskip3ex $H_0$\hfill
  &&  \hfill $1$\hfill
  &&  \hfill $-3$\hfill
  &\cr
  \+& \hskip4ex $K^a_{\mu}$\hfill
  &&  \hfill $2$\hfill
  &&  \hfill $-1$\hfill
  &&  \hskip3ex $(H_1)_{\mu}$\hfill
  &&  \hfill $1$\hfill
  &&  \hfill $-2$\hfill
  &\cr
  \+& \hskip4ex $L^a$\hfill
  &&  \hfill $2$\hfill
  &&  \hfill $-2$\hfill
  &&  \hskip3ex $(H_2)_{\mu\nu}$\hfill
  &&  \hfill $1$\hfill
  &&  \hfill $-1$\hfill
  &\cr
  \+& \hskip4ex $$\hfill
  &&  \hfill $$\hfill
  &&  \hfill $$\hfill
  &&  \hskip3ex $(H_3)_{\mu\nu\rho}$\hfill
  &&  \hfill $1$\hfill
  &&  \hfill $\kern7pt0$\hfill
  &\cr
}
\vskip1.0ex
\thicktablerule
}
$$
\vskip-0.0ex
\endinsert

The fields $s,\dbrs s$ and $\phi_k$ are such that
the short-distance singularities generated by
their insertion in correlation functions of the basic
fields $q,\ldots,\dbrs c$ are integrable at $D=4$.
Additional poles in $1/\eps$ are therefore excluded
when off-shell correlation functions are considered
and there is no difference in these cases between off- and
on-shell renormalization.

\subsection 4.3 Definition of the vertex functional

The complete list of source terms included in the functional integral
is thus
\equation{
  (J,q)+(\etabar,c)+(\cbar,\eta)+(K,\dbrs q)-(L,\dbrs c)
  \noenum
  \nexteq{1.0ex}
  {\phantom{(J,q)}}
  +(E,s)-(F,\dbrs s)
  +\sum_{k=0}^3(H_k,\phi_k),
  \enum
}
some relevant properties of the source fields being listed in table~1.

From the partition function $Z[B,J,\ldots,H_3]$ of the theory in presence
of the background gauge field and the source terms,
the generating functional
for the connected correlation functions of the fields
$q,\ldots,\phi_3$,
\equation{
  W[B,J,\ldots,H_3]=\ln(Z[B,J,\ldots,H_3]),
  \enum
}
is obtained as usual.
The Legendre transform
\equation{
  \Gamma[B,Q,\ldots,H_3]=W[B,J,\ldots,H_3]-(J,Q)-(\etabar,C)-(\Cbar,\eta),
  \enum
  \nexteq{2.0ex}
  Q^a_{\mu}(x)={\delta W\over\delta J^a_{\mu}(x)},
  \enum
  \nexteq{2.0ex}
  C^a(x)={\delta W\over\delta\etabar^a(x)},
  \qquad
  \Cbar^a(x)=-{\delta W\over\delta\eta^a(x)},
  \enum
}
in the source fields
for $q,c$ and $\cbar$
then leads to the vertex functional
$\Gamma[B,\ldots,H_3]$ of the theory,
all other source fields $E,\ldots,H_3$
and the background field
being spectators in this transformation.

\subsection 4.4 BRS symmetry

A little algebra now shows that the BRS symmetry implies
[\ref{ZJE}]
\equation{
  \int\rmd^D\!x\,\biggl\{
  {\delta\Gamma\over\delta Q_{\mu}^a}{\delta\Gamma\over\delta K_{\mu}^a}
  -{\delta\Gamma\over\delta C^a}{\delta\Gamma\over\delta L^a}
  +\lambda_0(D_{\mu}Q_{\mu})^a
  {\delta\Gamma\over\delta \Cbar^a}
  -E_{\mu}{\delta\Gamma\over\delta F_{\mu}}
  \noenum
  \nexteq{2.5ex}
  {\phantom{\int\rmd^D\!x\,\biggl\{}}
  -(\rmd^{\ast}\kern-1pt H_1){\delta \Gamma\over\delta H_0}
  +(\rmd^{\ast}\kern-1pt H_2)_{\mu}{\delta \Gamma\over\delta (H_1)_{\mu}}
  -(\rmd^{\ast}\kern-1pt H_3)_{\mu\nu}{\delta \Gamma\over\delta (H_2)_{\mu\nu}}
  \biggr\}=0.
  \enum
}
The first three terms in this equation are the usual ones deriving
from the BRS variation of the basic fields,
while all further terms reflect the transformation behaviour
of the added fields $s,\ldots,\phi_3$.

\subsection 4.5 Background gauge transformations

If the coloured source fields are transformed according to
\equation{
  \delta_{\omega}Q_{\mu}=[Q_{\mu},\omega],
  \enum
  \nexteq{2.0ex}
  \delta_{\omega}C=[C,\omega],
  \qquad
  \delta_{\omega}\Cbar=[\Cbar,\omega],
  \enum
  \nexteq{2.0ex}
  \delta_{\omega}K_{\mu}=[K_{\mu},\omega],
  \enum
  \nexteq{2.0ex}
  \delta_{\omega}L=[L,\omega],
  \enum
}
the vertex functional is invariant under
background gauge transformations up to an inhomogeneous term,
\equation{
  \delta_{\omega}\Gamma[B,\ldots,H_3]=\int\rmd^D\!x\,
  (H_3)_{\mu\nu\rho}\tr\{\partial_{\mu}\omega
  \partial_{\nu}B_{\rho}\},
  \enum
}
that derives from the non-invariance of $\phi_3$ [cf.~eq.~(3.20)].
In particular, the vertex functional is gauge invariant beyond the
tree level of perturbation theory.

\subsection 4.6 Shift symmetry

In the background gauge, the QCD action (2.1)
is a function of the gauge potential $A_{\mu}$, while
the gauge-fixing and the ghost action depend on both
the background and the quantum field.
Under an infinitesimal shift
\equation{
  \dshf B_{\mu}(x)=g_0\upsilon_{\mu}(x),
  \qquad
  \dshf q_{\mu}(x)=-\upsilon_{\mu}(x),
  \enum
}
of these fields by an arbitrary classical field
$\upsilon_{\mu}$, the total action transforms like
\equation{
  \dshf\Stot=\dbrs\left\{\int\rmd^D\!x\,\upsilon^a_{\mu}
  [(D_{\mu}+g_0\Ad q_{\mu})\cbar]^a\right\}.
  \enum
}
An identity used later, which derives from this property, is
\equation{
  {\delta\over\delta B^a_{\mu}}
  W[B,0,\ldots,0,\rmd^{\ast}\kern-1pt H_4]=0,
  \enum
}
where $(H_4)_{\mu\nu\rho\sigma}$ is any totally antisymmetric
tensor source field of rank $4$.

\section 5. Renormalization

In the following, the focus will be on the
vertex functions with either no or a single insertion
of the fields $s,\ldots,\phi_3$. The corresponding parts of the
vertex functional are denoted by $\Gzero$ and $\Gone$.
Clearly,
\equation{
  \Gzero[B,\ldots,L]=\Gamma[B,\ldots,L,0,\ldots,0],
  \enum
}
while $\Gone[B,\ldots,H_3]$ coincides with the part of
the vertex functional that depends linearly on the source
fields $E,\ldots,H_3$.
The discussion in this section largely follows the one in
ref.~[\ref{BFM}], where the renormalizability of $\Gzero[B,\ldots,L]$
was proved. Here the goal is
to extend this result to $\Gone[B,\ldots,H_3]$.

\subsection 5.1 Renormalized vertex functional

The vertex functional is tentatively renormalized by
scaling the source fields,
\equation{
  \GR[B,Q,\Cbar,C,K,L,E,\ldots,H_3]
  \noenum
  \nexteq{2.0ex}
  \qquad=
  \Gamma[B,Z_3^{1/2}Q,\widetilde{Z}_3^{1/2}\Cbar,\widetilde{Z}_3^{1/2}C,
  \widetilde{Z}_3^{1/2}K,Z_3^{1/2}L,
  \noenum
  \nexteq{2.0ex}
  {\phantom{\qquad=\Gamma[}}
  Z_E(E+X_E\rmd^{\ast}\kern-1pt H_2),Z_F(F-X_EH_1),
  Z_{H_0}H_0,\ldots,Z_{H_3}H_3],
  \enum
}
and by expressing the bare coupling and gauge parameter through
the renormalized coupling $g$ and gauge parameter $\lambda$
according to
\equation{
  g_0=\mu^{\eps}Z_1Z_3^{-3/2}g,\qquad\hbox{$\mu$: normalization mass,}
  \enum
  \nexteq{2.0ex}
  \lambda_0=Z_3^{-1}\lambda.
  \enum
}
In these equations, $Z_1,Z_3$ and $\widetilde{Z}_3$ are the renormalization
constants already required for the renormalization of the
theory without insertions of the fields $s,\ldots,\phi_3$.
Some of the other renormalization constants are not independent
and satisfy
\equation{
  Z_F=Z_E(Z_3\widetilde{Z}_3)^{1/2},
  \enum
  \nexteq{2.0ex}
  Z_{H_k}=Z_H(Z_3\widetilde{Z}_3)^{(3-k)/2},\quad k=0,\ldots,3.
  \enum
}
The additive renormalizations proportional to $X_E$
are included in eq.~(5.2), because the
field $\phi_1$ mixes with $\dbrs s$ and $\phi_2$ with $\rmd s$.
A non-zero mixing actually already occurs
at one-loop order of perturbation theory.
In the following, minimal subtraction (i.e.~the MS scheme) is assumed
for all renormalization constants.

Equations (5.2)--(5.6) are such that
the renormalization preserves
the form of the BRS identity (4.13), viz.
\equation{
  \int\rmd^D\!x\,\biggl\{
  {\delta\GR\over\delta Q_{\mu}^a}{\delta\GR\over\delta K_{\mu}^a}
  -{\delta\GR\over\delta C^a}{\delta\GR\over\delta L^a}
  +\lambda(D_{\mu}Q_{\mu})^a
  {\delta\GR\over\delta \Cbar^a}
  -E_{\mu}{\delta\GR\over\delta F_{\mu}}
  \noenum
  \nexteq{2.5ex}
  {\phantom{\int\rmd^D\!x\,\biggl\{}}
  -(\rmd^{\ast}\kern-1pt H_1){\delta \GR\over\delta H_0}
  +(\rmd^{\ast}\kern-1pt H_2)_{\mu}{\delta \GR\over\delta (H_1)_{\mu}}
  -(\rmd^{\ast}\kern-1pt H_3)_{\mu\nu}{\delta \GR\over\delta (H_2)_{\mu\nu}}
  \biggr\}=0.
  \enum
}
Moreover, the shift-symmetry identity (4.21) continues to hold
when $W[B,\ldots,H_3]$
is replaced by the generating functional $W_R[B,\ldots,H_3]$
of the renormalized connected correlation functions.
In the case of the background gauge symmetry,
the trans\-formation law for the renormalized
vertex functional,
\equation{
  \delta_{\omega}\GR[B,\ldots,H_3]=Z_H\int\rmd^D\!x\,
  (H_3)_{\mu\nu\rho}\tr\{\partial_{\mu}\omega
  \partial_{\nu}B_{\rho}\},
  \enum
}
however involves the renormalization constant $Z_H$,
which already shows that $Z_H$ cannot diverge
in the limit $D\to4$ if $\GRone[B,\ldots,H_3]$ is finite.

\subsection 5.2 Loop expansion of the renormalized vertex functional

The renormalized vertex functional may be expanded in a series
\equation{
  \GR=\sum_{l=0}^{\infty}\GRn{l}
  \enum
}
of terms of increasing loop order $l$, the lowest-order term being
\equation{
  \GRn{0}=-\hat{S}_{\rm tot}
  \noenum
  \nexteq{0.5ex}
  {\phantom{\GRn{0}=}}
  +(K,\dbrsc Q)-(L,\dbrsc C)
  +(E,\hat{s})-(F,\dbrsc\hat{s})+\sum_{k=0}^3(H_k,\hat{\phi}_k).
  \enum
}
All hatted fields in this formula and the
hatted total action
are obtained from the corresponding expressions in the quantum fields
by substituting
$q_{\mu}\to Q_{\mu}$, $c\to C$, $\cbar\to\Cbar$, $g_0\to\mu^{\eps}g$
and $\lambda_0\to\lambda$.
The BRS variation
\equation{
  \dbrsc Q_{\mu}=
  (D_{\mu}+\mu^{\eps}g\Ad Q_{\mu})C,
  \enum
  \nexteq{2.0ex}
  \dbrsc C=
  -\mu^{\eps}gC^2,
  \enum
}
acting on the source fields must be distinguished from the one acting
on the quantum fields, but has identical algebraic properties.

\subsection 5.3 Proof of finiteness: first steps

The proof of finiteness of $\GRone[B,\ldots,H_3]$ and thus of $Z_H$
proceeds by induction over the loop order $l$. At a given order $n$,
the induction hypothesis is that the divergences of
$\GRonen{l}$ can be canceled at all orders $l<n$
by setting $Z_H=1$ and by adjusting the $l$-loop coefficients of
$Z_E$ and $X_E$. The task is then to show that the same is the case
at loop order $n$.

First this requires the structure of the divergent part
$\Delta\GRonen{n}$ of $\GRonen{n}$ to be determined for vanishing
$n$-loop terms $Z_{E,n},X_{E,n},Z_{H,n}$
of $Z_E,X_E$ and $Z_H$,
their contribution to the vertex functional
at this order,
\equation{
  Z_{E,n}\{(E,\hat{s})-(F,\dbrsc\hat{s})\}
  \noenum
  \nexteq{1.0ex}
  \qquad
  +
  X_{E,n}\{(\rmd^{\ast}\kern-1pt H_2,\hat{s})+(H_1,\dbrsc\hat{s})\}
  +
  Z_{H,n}\sum_{k=0}^3(H_k,\hat{\phi}_k),
  \enum
}
being taken into account in subsect.~5.5.

General principles imply that
\equation{
  \Delta\GRonen{n}=\int\rmd^D\! x\,p(x),
  \enum
}
where $p(x)$ is a local polynomial in the source
fields $B,Q,\ldots,H_3$ and their derivatives,
which must have dimension $4$, ghost number $0$ and be
linear in $E,\ldots,H_3$. Partial integration
moreover allows any terms with derivatives of these latter fields to be
traded for terms in which they appear without derivatives.
The field $p(x)$ then inherits the invariance of
$\Delta\GRonen{n}$ under Lorentz and background gauge transformations
(since $Z_{H,n}$ is, at this point, set to zero).

All these properties already strongly constrain the form of $p(x)$.
Recalling table~1,
inspection shows that the field cannot depend
on the fields $\Cbar$, $K$ or $L$.
Moreover, the terms in $p(x)$ depending on the fields
$E$ and $H_0$ must be proportional to
$E_{\mu}(x)\hat{s}_{\mu}(x)$ and $H_0(x)\hat{\phi}_0(x)$.

\subsection 5.4 Consequences of the BRS symmetry

Further constraints on $\Delta\GRonen{n}$ derive from
the BRS identity (5.7), which holds at all
loop orders and all orders in the source fields $E,\ldots,H_3$.
At loop order $n$,
and for the terms linear in these fields,
the identity together with the induction hypothesis and the
leading-order form (5.10) of
the vertex functional implies
\equation{
  \int\rmd^D\!x\,\biggl\{
  \dbrsc Q_{\mu}^a
  {\delta\Delta\GRonen{n}\over\delta Q_{\mu}^a}
  +\dbrsc C^a{\delta\Delta\GRonen{n}\over\delta C^a}
  -E_{\mu}{\delta\Delta\GRonen{n}\over\delta F_{\mu}}
  -(\rmd^{\ast}\kern-1pt H_1)
  {\delta \Delta\GRonen{n}\over\delta H_0}
  \noenum
  \nexteq{2.5ex}
  {\phantom{\int\rmd^D\!x\,\biggl\{}}
  +(\rmd^{\ast}\kern-1pt H_2)_{\mu}
  {\delta\Delta\GRonen{n}\over\delta (H_1)_{\mu}}
  -(\rmd^{\ast}\kern-1pt H_3)_{\mu\nu}
  {\delta\Delta\GRonen{n}\over\delta (H_2)_{\mu\nu}}
  \biggr\}=0.
  \enum
}
If only the first two terms were present, the left-hand side
of this equation would coincide with $\dbrsc\Delta\GRonen{n}$.
The equation thus relates the BRS variation
of the terms proportional to $E,H_1,H_2$ and $H_3$
to the terms  proportional to $F,H_0,H_1$ and $H_2$.
As a consequence,
\equation{
  \Delta\GRonen{n}=
  z_E\bigl\{(E,\hat{s})-(F,\dbrsc\hat{s})\bigr\}+\sum_{k=0}^3(H_k,\hat{f}_k),
  \enum
}
where $z_E$ is a (divergent) constant
and $\hat{f}_k$, $k=0,\ldots,3$, some gauge-invariant forms of rank $k$,
with dimension $3$ and ghost number $3-k$, satisfying
\equation{
  \dbrsc \hat{f}_k=\rmd \hat{f}_{k-1}
  \enum
}
for all $k=1,2,3$.

The discussion
in appendix A of the descent equations for quantum fields
carries over literally to the case of the descent equations (5.17)
and shows that these equations have
only two linearly independent solutions with the required
properties.
As a result,
\equation{
  \Delta\GRonen{n}=
  (z_EE+x_E\rmd^{\ast}\kern-1pt H_2,\hat{s})
  -(z_EF-x_EH_1,\dbrsc\hat{s})
  \noenum
  \nexteq{2.0ex}
  {\phantom{\Delta\GRonen{n}=}}
  +z_H\sum_{k=0}^3(H_k,\hat{\phi}_k)
  -z_H(H_3,\hat{\phi}_3)_{g=0},
  \enum
}
where $x_E$ and $z_H$ are further (divergent) coefficients.

\subsection 5.5 Proof of finiteness: final steps

Now when the counterterms (5.13) are included in the
vertex functional,
all terms on the right of eq.~(5.18) except for the last one
can be canceled by adjusting the $n$-loop coefficients of
$Z_E,X_E$ and $Z_H$.
Since the uncanceled term only depends on $B$ and $H_3$,
it is a spectator in the Legendre transform that leads
from the renormalized vertex functional to
the generating functional $W_R[B,\ldots,H_3]$
of the renormalized correlation functions.
The latter is therefore finite too at $n$-loop order
apart from this additive divergent term
and terms of higher than linear order in the
source fields $E,\ldots,H_3$.

Such a divergent term is however excluded by
the shift-symmetry relation (4.21) (with $W\to W_R$)
and its coefficient $z_H$ must hence be equal to zero.
The terms in eq.~(5.18) proportional to $z_H$ are thus
absent and all divergences at $n$-loop order can be
canceled by setting $Z_H=1$ and adjusting $Z_E$ and $X_E$,
as was to be shown.

\subsection 5.6 Inclusion of the quark fields

In presence of the quark fields, $\phi_3$
requires an additive renormalization proportional to
the flavour-singlet axial current, which is here represented by
the tensor field
\equation{
  A^s_{\mu\nu\rho}(x)=\sum_{r=1}^{\Nf}
  \psibar_r(x)\dirac{[\mu}\dirac{\vphantom{[}\nu}\dirac{\rho]}\psi_r(x).
  \enum
}
After adding
source terms for the quark and antiquark fields,
their BRS variation and the axial current (5.19),
the finiteness of $\GRone$ can then again be proved following the
steps taken in the case of the pure gauge theory.

Since $A^s$ is invariant under both the BRS
and the background gauge symmetry,
there is now a third solution,
$\hat{f}_k=\delta_{k3}\hat{A}^s$, of the descent equations (5.17)
with all the required properties.
The mixing of $\phi_3$ with $A^s$ derives from the existence
of this additional solution, but a
multiplicative renormalization of $\phi_3$ remains excluded.

There is, on the other hand, no field that could mix with
the axial current. The results obtained in appendix B in fact
show that no BRS and gauge invariant 3-form
of dimension $3$ can be built from the gauge and ghost
fields alone. The Lorentz and flavour symmetry then imply
that the current must renormalize multiplicatively.

\section 6. Flavour-singlet axial Ward identity

Since the axial anomaly does not require multiplicative
renormalization, the relation between the bare
and the renormalized fields that appear in the flavour-singlet
axial-current conservation equation is slightly simplified.
The structure of the equation in the renormalized theory
in four dimensions is then easily determined,
but there is little new here and the section
is included mainly for completeness.
All statements made in the following refer to standard QCD
with vanishing background field.

\subsection 6.1 Renormalized fields

The renormalized fields participating in the Ward identity are
\equation{
  \FFR=\frac{1}{2}\eps_{\mu\nu\rho\sigma}
  \left\{F^a_{[\mu\nu}F^a_{\rho\sigma]}
  +\frac{1}{3}
  Z_{FA}\partial_{[\mu}\As{\nu\rho\sigma]}\right\},
  \enum
  \nexteq{3.0ex}
  \AsR{\mu}=\frac{1}{6}\eps_{\mu\nu\rho\sigma}Z_A\As{\nu\rho\sigma},
  \enum
  \nexteq{1.0ex}
  \mPR=
  \frac{1}{24}\eps_{\mu\nu\rho\sigma}
  Z_mZ_P\sum_{r=1}^{\Nf}
  \mb{r}P^{rr}_{\mu\nu\rho\sigma}.
  \enum
}
In these equations,
the renormalization constants $Z_m,Z_A$ and $Z_P$
are for the quark masses and for the flavour-singlet axial current and
density, while $Z_{FA}=\rmO(g^4)$ is a mixing
coefficient. Minimal subtraction is assumed, as before,
and the contraction with the Levi--Civita symbol is performed
only after passing to $D=4$ dimensions.

The anomalous dimensions of $\AsR{\mu}$ and $\mPR$ are
\equation{
  \gamma_A=(-\eps g+\beta){\partial\ln Z_A\over\partial g},
  \enum
  \nexteq{2.0ex}
  \gamma_{mP}=(-\eps g+\beta){\partial\ln(Z_mZ_P)\over\partial g}
  =\gamma_m+\gamma_P,
  \enum
}
where $\beta=-b_0g^3-b_1g^5+\ldots$ denotes the $\beta$-function at $\eps=0$.
Since $\FFR$ mixes with $\partial_{\mu}\AsR{\mu}$,
the associated anomalous dimension is a $2\times 2$ matrix,
\equation{
  \gamma_F=\pmatrix{0 & \gamma_{FA}\cr
                  \noalign{\vskip1.5ex}
                  0 & \gamma_A \cr},
  \qquad
  \gamma_{FA}=(-\eps g+\beta){\partial Z_{FA}\over\partial g}Z_A^{-1},
  \enum
}
acting on these fields.

\subsection 6.2 Renormalization-group-invariant (RGI) fields

RGI fields are related to the renormalized ones through
finite renormalization factors chosen such that the anomalous dimensions
vanish. In the case of a multiplet $\OR{k}$, $k=1,\ldots,n$,
of fields with anomalous-dimension matrix $\gamma$,
the RGI fields are given by
\equation{
  \ORGI{k}=\sum_{l=1}^n\Rmat_{kl}\OR{l},
  \enum
}
where $\Rmat$ is an $n\times n$ matrix
satisfying
\equation{
  \beta{\partial\Rmat\over\partial g}+\Rmat\gamma=0
  \enum
}
plus some conventional boundary condition at $g=0$.
Apart from having vanishing anomalous dimension,
RGI fields are independent of the renormalization scheme
and any relations among them are therefore universally valid.

In the case of the fields considered here, the boundary condition
$\lim_{g\to0}\Rmat=1$
can be imposed and the RGI fields are then given by
\equation{
  \FFRGI=\FFR+\Xmat_{FA}\partial_{\mu}\AsR{\mu},
  \enum
  \nexteq{2.0ex}
  \AsRGI{\mu}=\Xmat_A\AsR{\mu},
  \enum
  \nexteq{2.0ex}
  \mPRGI=\Xmat_{mP}\mPR,
  \enum
}
where
\equation{
  \Xmat_O=
  \exp\biggl\{-\int_0^g\rmd h\,{\gamma_{O}(h)\over\beta(h)}\biggr\},
  \qquad
  O=A,mP,
  \enum
  \nexteq{3.0ex}
  \Xmat_{FA}=-\Xmat_A\int_0^g\rmd h\,
  {\gamma_{FA}(h)\over\beta(h)\Xmat_A(h)}
  \enum
}
(the integrals are all absolutely convergent,
since the anomalous dimensions $\gamma_A$, $\gamma_{mP}$ and
$\gamma_{FA}$ are of order $g^4$).
The factor
$\Xmat_{mP}$ can, incidentally, also be determined by
matching the normalizations of the axial and scalar quark densities
as in ref.~[\ref{Larin}], for example.

\subsection 6.3 Ward identity

In terms of the RGI fields,
and for any product $\obs$ of fields at non-zero distances from $x$,
the flavour-singlet Ward identity assumes the form
\equation{
  \left\langle\{\partial_{\mu}\AsRGI{\mu}(x)
  +k_1\mPRGI(x)+k_2\FFRGI(x)\}\obs\right\rangle=0.
  \enum
}
Since the renormalization group excludes
a dependence of the
coefficients $k_1$ and $k_2$ on the gauge coupling,
their values
\equation{
  k_1=-2,
  \qquad
  k_2={\Nf\over16\pi^2},
  \enum
}
coincide with the ones obtained at 1-loop order of
perturbation theory.

If the Ward identity is written in terms of the minimally subtracted
field $\FFR$ instead of $\FFRGI$,
as in refs.~[\ref{Larin}--\ref{AhmedII}],
the equation becomes
\equation{
  \left\langle\{\partial_{\mu}\AsRp{\mu}(x)
  +k_1\mPRGI(x)+k_2\FFR(x)\}\obs\right\rangle=0,
  \enum
  \nexteq{2.0ex}
  \AsRp{\mu}=(\Xmat_A+k_2\Xmat_{FA})\AsR{\mu}.
  \enum
}
In this renormalization scheme,
the anomalous dimensions satisfy
\equation{
  \gamma_A'=-k_2\gamma_{FA}',
  \enum
}
as already noted in ref.~[\ref{AhmedII}],
and the finite renormalization factor $\Xmat_A+k_2\Xmat_{FA}$
can be computed by requiring (6.16) to hold in the massless
theory [\ref{Larin}].

\section 7. Concluding remarks

The fact that the topological charge density
does not require multiplicative renormalization
derives from its algebraic properties,
namely that it coincides with the
exterior differential of a gauge-variant local 3-form,
the Chern--Simons form.
Eventually the normalization of the density
is fixed by the inhomogeneous gauge transformation
behaviour of the latter.

A straightforward argumentation along this line is
however not possible in perturbation theory
in view of the required gauge fixing.
Use had instead to be made of the
background gauge and the BRS symmetry,
whose application to the Chern--Simons form generates
a chain of forms with increasing ghost number.
All these forms must be included in the renormalization
process and a multiplicative renormalization
of the Chern--Simons form (and thus of the charge density)
is then seen to be excluded by the symmetries of the
QCD vertex functional. The other forms
however require multiplicative renormalization and some
additive renormalization as well.

Specific renormalization properties like the one discussed in
this paper can depend on the chosen regularization
of the theory. Simple expressions for the topological
charge density in lattice QCD, for example, need not be
exactly representable through a discrete version of the
Chern--Simons form and may consequently require multiplicative
renormalization. After renormalization and removal of the
regularization,
the RGI form (6.14) of the flavour-singlet chiral Ward identity
however holds in these cases too.

\appendix A. Notation

\vskip-2.5ex

\subsection A.1 Gauge group

The Lie algebra of the gauge group $\SUn$
may be identified with the space of all complex antihermitian
$N\times N$ matrices with vanishing trace.
If $T^a$, $a=1,\ldots,N^2-1$, is a basis of such matrices
satisfying
\equation{
  \tr\{T^aT^b\}=-\frac{1}{2}\delta^{ab},
  \enum
}
the general element $X$ of the Lie algebra is given by
$X=X^aT^a$ with real components $X^a=-2\kern1pt\tr\{XT^a\}$
(repeated indices are automatically summed over).

While the quark fields are assumed to be in the fundamental
representation of the gauge group, the gauge and ghost fields take values
in its Lie algebra. The adjoint action of the latter on itself
is defined by
\equation{
  \Ad X\cdot Y=[X,Y]=f^{abc}X^aY^bT^c,
  \enum
}
where $f^{abc}$ are the
$\SUn$ structure constants in the chosen basis of group generators.

\subsection A.2 Dimensional regularization

The theory is defined in the standard manner
in $D=4-2\eps$ Euclidean dimensions. Lorentz indices
run from $0$ to $3$ in $D=4$ dimensions and formally to $D-1$ in arbitrary
dimensions, i.e.~the trace of the Kronecker delta $\delta_{\mu\nu}$ is
equal to $D$.

The Dirac matrices $\dirac{\mu}$ in $D$ dimensions are formal objects
satisfying
\equation{
  \{\dirac{\mu},\dirac{\nu}\}=2\delta_{\mu\nu}.
  \enum
}
By taking products and linear combinations, the Dirac matrices
generate an infinite dimensional linear space.
The trace $\tr\{\cdot\}$ is a mapping from this space to
the space of polynomials in Kronecker deltas, which is
implicitly defined by its linearity and cyclicity,
the normalization convention
\equation{
  \tr\{1\}=4,
  \enum
}
the Dirac algebra (A.3)
and the rule that products of
odd numbers of Dirac matrices have vanishing trace.

In $D=4$ dimensions, the Dirac matrices are assumed to be Hermitian
and the fifth Dirac matrix is taken to be
\equation{
  \dirac{5}=\dirac{0}\dirac{1}\dirac{2}\dirac{3},
  \enum
}
but no attempt is made to assign a meaning to $\dirac{5}$
in arbitrary dimensions. The same applies to the Levi--Civita symbol
$\eps_{\mu\nu\rho\sigma}$, which is normalized such that
$\eps_{0123}=1$.

\subsection A.3 Differential forms

Differential forms $f(x)$ of rank $n$ are homogeneous polynomials
\equation{
  f(x)=f(x)_{\mu_1\ldots\mu_n}\dx{\mu_1}\ldots\dx{\mu_n}
  \enum
}
in the Grassmann algebra generated by the anticommuting
symbols $\dx{\mu}$. The coefficients $f(x)_{\mu_1\ldots\mu_n}$
may be real, complex or take values in the Lie algebra of $\SUn$,
for example.

The exterior differential $\rmd$ acts on such forms according to
\equation{
  \rmd f(x)=\dx{\mu}\partial_{\mu}f(x).
  \enum
}
Clearly, $\rmd^2=0$ and
\equation{
  \rmd(f(x)g(x))=\rmd f(x)g(x)+(-1)^nf(x)\rmd g(x)
  \enum
}
if $f(x)$ has rank $n$.

\appendix B. Solution of the descent equations

The goal in this appendix is to find the general solution
of the descent equations
\equation{
  \dbrs f_k=\rmd f_{k-1},
  \quad k=1,2,3,
  \enum
}
in the space of local gauge-invariant forms
$f_k$ of rank $k=0,\ldots,3$,
with ghost number $3-k$ and dimension 3, which
can be composed from
the fields $B_{\mu}^a,q_{\mu}^a,c^a$ and their derivatives.

\subsection B.1 Gauge-covariant exterior differential

Let $f$ be any differential form of rank $n$
with values in the space of $N\times N$ matrices
of elements of a complex Grassmann algebra.
The gauge-covariant exterior differential
$\rmd^B\!$ acts on $f$ according to
\equation{
  \dB f=\rmd f+Bf+(-1)^{n+1}fB.
  \enum
}
If $f$ is a gauge-covariant expression in the basic fields, i.e.~if
\equation{
  \delta_{\omega}f=[f,\omega],
  \enum
}
its differential $\dB f$ is gauge-covariant too.
Moreover,
\equation{
  \rmd^B(\dB f)=[G,f],\qquad
  G=\rmd B+B^2,
  \enum
  \nexteq{2.0ex}
  \dB G=0,
  \enum
  \nexteq{2.0ex}
  \dB(fg)=\dB fg+(-1)^nf\dB g,
  \enum
  \nexteq{2.0ex}
  \rmd\left(\tr\{f\}\right)=\tr\{\dB f\}.
  \enum
}
Using these rules, the differential of the trace of any gauge-covariant
polynomial in the basic fields can be worked out and yields
expressions of the same type.

\subsection B.2 Gauge-invariant forms

The dimension, ghost number and gauge invariance of the differential
forms $f_k$
implies that they are linear combinations of terms of the form $\tr\{\obs\}$,
where $\obs$ is a product of three of the fields $B_{\mu},q_{\mu},c$
or of one of these fields and another one with a derivative acting on it.
Moreover, none of the Lorentz indices of the fields may be contracted,
i.e.~$\obs$ must be a product of $B,q,c,\rmd B,\rmd q$ and $\rmd c$.
The gauge-invariant terms with ghost number $0,\ldots,3$,
\equation{
  h_1=\tr\{q^3\},
  \qquad
  h_2=\tr\{q\kern1pt\dB q\},
  \qquad
  h_3=\tr\{qG\},
  \enum
  \nexteq{2.0ex}
  h_4=\tr\{cq^2\},
  \qquad
  h_5=\tr\{c\kern1pt\dB q\},
  \qquad
  h_6=\tr\{\dB cq\},
  \qquad
  h_7=\tr\{cG\},
  \enum
  \nexteq{2.0ex}
  h_8=\tr\{c^2q\},
  \qquad
  h_9=\tr\{c\kern1pt\dB c\},
  \enum
  \nexteq{2.0ex}
  h_{10}=\tr\{c^3\},
  \enum
}
are then easily found by applying a gauge variation
to the general linear combination of all possible terms.

\subsection B.3 General solution of the descent equations

Recalling the discussion in subsect.~3.3,
the sequence of forms
\equation{
  f_k=\phi_k-\delta_{k3}\left.\phi_3\right|_{g_0=0},
  \quad k=0,\ldots,3,
  \enum
}
is easily shown to satisfy the descent equations (B.1) and
all other requirements too.
An obvious second solution is
\equation{
  f_k=\delta_{k1}\dbrs s+\delta_{k2}\rmd s,
  \qquad
  s=\tr\{cq\},
  \enum
}
but there are no further linearly independent solutions.

The proof of this statement begins by noting that the forms
$f_k$ must be linear combinations
\equation{
  f_3=\sum_{k=1}^3c_kh_k,\quad f_2=\sum_{k=4}^7c_kh_k,\quad\ldots
  \enum
}
of the forms $h_1,\ldots,h_{10}$ with some coefficients
$c_1,\ldots,c_{10}$. Since
\equation{
  \phi_3-\left.\phi_3\right|_{g_0=0}=
  \frac{2}{3}g_0^3h_1+g_0^2h_2+2g_0h_3,
  \enum
  \nexteq{2.0ex}
  \rmd s=h_5+h_6,
  \enum
}
the coefficients $c_3$ and $c_5$
can be nullified by subtracting
a linear combination of the solutions (B.12),(B.13)
and it remains to be shown
that the descent equations imply the
vanishing of all coefficients $c_1,\ldots,c_{10}$ if $c_3=c_5=0$.

Noting
\equation{
  \dbrs q=\dB c+g_0[q,c],
  \enum
  \nexteq{2.0ex}
  \dbrs\dB q=[G,c]+g_0[\dB q,c]-g_0\{\dB c,q\},
  \enum
}
some algebra yields
\equation{
  \dbrs h_1=3\kern1pt\tr\{\dB cq^2\},
  \enum
  \nexteq{2.0ex}
  \dbrs h_2=\tr\{\dB c\kern1pt\dB q-2g_0\dB cq^2-[q,c]G\},
  \enum
}
for the forms with ghost number 0, while
\equation{
  \rmd h_4=\tr\{\dB c q^2+[q,c]\dB q\},
  \enum
  \nexteq{2.0ex}
  \rmd h_6=-\tr\{\dB c\dB q+[q,c]G\},
  \enum
  \nexteq{2.0ex}
  \rmd h_7=\tr\{\dB cG\}.
  \enum
}
The matching of
the independent terms on the two sides of the equation
$\dbrs f_3=\rmd f_2$ then shows that
$c_1=c_2=c_4=c_6=c_7=0$ and thus $f_2=f_3=0$.
Finally, since the forms
\equation{
  \rmd h_8=\tr\{c^2\dB q-[q,c]\dB c\},
  \enum
  \nexteq{2.0ex}
  \rmd h_9=\tr\{\dB c\kern1pt\dB c-2c^2G\},
  \enum
  \nexteq{2.0ex}
  \rmd h_{10}=3\kern1pt\tr\{c^2\dB c\},
  \enum
}
are linearly independent,
the remaining coefficients $c_8,c_9,c_{10}$ must vanish too.

\beginbibliography


\bibitem{VermaserenLarin}
S. A. Larin, J. A. M. Vermaseren,
{\it The $\alpha_s^3$ corrections to the Bjorken sum rule for
polarized electroproduction and to the Gross-Llewellyn Smith
sum rule},
Phys. Lett. B259 (1991) 345

\bibitem{Larin}
S. A. Larin,
{\it The renormalization of the axial anomaly in dimensional regularization},
Phys. Lett. B303 (1993) 113

\bibitem{Zoller}
M. Zoller,
{\it OPE of the pseudoscalar gluonium correlator in massless QCD
to three-loop order},
JHEP 07 (2013) 40

\bibitem{AhmedI}
T. Ahmed, T. Gehrmann, P. Mathews, N. Rana, V. Ravindran,
{\it Pseudo-scalar form factors at three loops in QCD},
JHEP 11 (2015) 169

\bibitem{AhmedII}
T. Ahmed, L. Chen, M. Czakon,
{\it Renormalization of the flavor-singlet axial-vector current and its
anomaly in dimensional regularization},
JHEP 05 (2021) 087


\bibitem{AB}
S. L. Adler, W. A. Bardeen,
{\it Absence of higher-order corrections in the
anomalous axial-vector divergence equation},
Phys. Rev. 182 (1969) 1517


\bibitem{BMS}
P. Breitenlohner, D. Maison, K. S. Stelle,
{\it Anomalous dimensions and the Adler\--Bardeen theorem in
supersymmetric Yang--Mills theories},
Phys. Lett. B134 (1984) 63


\bibitem{BRSI}
C. Becchi, A. Rouet, R. Stora,
{\it Renormalization of the Abelian Higgs--Kibble model},
Commun. Math. Phys. 42 (1975) 127

\bibitem{BRSII}
C. Becchi, A. Rouet, R. Stora,
{\it Renormalization of gauge theories},
Ann. Phys. (NY) 98 (1976) 287


\bibitem{ZJE}
J. Zinn-Justin,
{\it Renormalization of gauge theories},
in: Trends in Elementary Particle Theory,
H. Rollnik, K. Dietz (eds.),
Lecture Notes in Physics 37 (Springer-Verlag, Berlin, 1975)


\bibitem{DeWittI}
B. S. DeWitt,
{\it Quantum theory of gravity. 2. The manifestly covariant theory},
Phys. Rev. 162 (1967) 1195

\bibitem{DeWittII}
B. S. DeWitt,
{\it Quantum theory of gravity. 3. Applications of the covariant theory},
Phys. Rev. 162 (1967) 1239

\bibitem{KlubergZuberI}
H. Kluberg-Stern, J. B. Zuber,
{\it Renormalization of non-Abelian gauge theories in a background
field gauge. 1. Green functions},
Phys. Rev. D12 (1975) 482

\bibitem{KlubergZuberII}
H. Kluberg-Stern, J. B. Zuber,
{\it Renormalization of non-Abelian gauge theories in a background
field gauge. 2. Gauge-invariant operators},
Phys. Rev. D12 (1975) 3159


\bibitem{StoraI}
R. Stora,
{\it Continuum gauge theories}, in:
New Developments in Quantum Field Theory and Statistical Mechanics
(Carg\`ese 1976),
M. L\'evy, P. Mitter (eds.)
(Plenum Press, New York, 1977)

\bibitem{StoraII}
R. Stora,
{\it Algebraic structure and topological origin of anomalies}, in:
Progress in Gauge Field Theory (Carg\`ese 1983),
G.'t Hooft et al. (eds.)
(Plenum Press, New York, 1984)

\bibitem{Zumino}
B. Zumino,
{\it Chiral anomalies and differential geometry}, in:
Relativity, Groups and Topology II (Les Houches 1983),
B. S. DeWitt, R. Stora (eds.)
(North Holland, Amsterdam, 1984)


\bibitem{BFM}
M. L\"uscher, P. Weisz,
{\it Background field technique and renormalization in lattice gauge theory},
Nucl. Phys. B452 (1995) 213


\bibitem{Bertlmann}
R. A. Bertlmann,
{\it Anomalies in quantum field theory}
(Clarendon Press, Oxford, 2000)

\endbibliography

\bye